\documentclass[aps,preprint,superscriptaddress,showpacs,showkeys]{revtex4}%
\usepackage{amsfonts}
\usepackage{amsmath}
\usepackage{amssymb}
\usepackage{graphicx}%
\begin{document}
\title{Destabilization of Neutron Stars by Type I Dimension Bubbles\\ }
\author{E.I. Guendelman}
\affiliation{Department of Physics, Ben-Gurion University of the Negev, Box 653, IL-84105
Beer Sheva, Israel}
\email{guendel@bgumail.bgu.ac.il}
\author{J.R. Morris}
\affiliation{Physics Dept., Indiana University Northwest, 3400 Broadway, Gary, Indiana
46408, USA}
\email{jmorris@iun.edu}

\begin{abstract}
An inhomogeneous compactification of a higher dimensional spacetime can result
in the formation of type I dimension bubbles, i.e., nontopological solitons
which tend to absorb and entrap massive particle modes. We consider possible
consequences of a neutron star that harbors such a soliton. The astrophysical
outcome depends upon the model parameters for the dimension bubble, with a
special sensitivity to the bubble's energy scale. For relatively small energy
scales, the bubble tends to rapidly consume the star without forming a black
hole. For larger energy scales, the bubble grows to a critical mass, then
forms a black hole within the star, which subsequently causes the remaining
star to collapse. It is possible that the latter scenario is associated with
core collapse explosions and gamma ray bursts.

\end{abstract}

\maketitle

\section{Introduction}

An inhomogeneous compactification of a higher dimensional spacetime with
compact extra dimensions may result in the production of dimension
bubbles\cite{BG}-\cite{JM1} inhabiting the effective 4d spacetime. A dimension
bubble may be either of the type I or type II variety, depending upon the
parameters of the effective potential\cite{JM1}. From a 4d perspective, a type
I(II) dimension bubble is a nontopological soliton characterized by a scale
factor for the extra dimensions which becomes much larger (smaller) inside the
bubble than outside. For example, if the characteristic size of the extra
dimensions varies from a Planck size to a TeV$^{-1}$ size across the bubble
wall, the scale factor can change by roughly 16 orders of magnitude with the
size of the extra dimension remaining everywhere microscopic. Massive fermions
and bosons in the effective 4d theory have masses varying inversely with the
extra dimensional scale factor\cite{JM1}, so that they tend to become trapped
inside a type I bubble, as in the case with the\ nontopological soliton of the
type studied by Frieman, Gleiser, Gelmini, and Kolb\cite{NTS}. Associated with
the rapidly varying extra dimensional scale factor within a type I bubble wall
is a very strong, short ranged attractive force. Therefore, massive particles
that are incident upon a type I bubble tend to be absorbed by it. As energy is
absorbed by the bubble, its mass and size both increase.

An interesting astrophysical situation can emerge if a type I dimension bubble
is situated within an environment of high particle density, such as the core
of a neutron star. As with the case of a Q-ball lodged within a neutron
star\cite{Q1,Q2}, the soliton may consume the star or form a black hole which
then consumes the star. However, the solitonic properties, particle absorption
rates, and energy absorption processes for dimension bubbles are quite
different from those for Q-balls, and we focus attention here upon the
consumption of a neutron star by a type I dimension bubble. We begin with an
expression for the mass absorption rate $\dot{M}$ for a type I dimension
bubble at the center of a neutron star, in terms of the Fermi momentum $p_{F}$
and the bubble radius $R$. The Fermi momentum can be related to the neutron
number density $n$, so that we obtain a simple approximate proportionality
between $\dot{M}$ and $R^{2}$. For a bubble that rapidly adjusts its radius so
that it is approximately in mechanical equilibrium, the bubble mass $M$ is
proportional to $R^{2}$, and $\dot{M}/M$ becomes a constant which depends upon
the bubble wall tension $\sigma$. Different scenarios can ensue, depending
upon the value of $\sigma$.

\section{Spherical Accretion of Neutron Star by Dimension Bubble}

For convenience we take a simplistic view of a neutron star as a ball of
neutrons with typical mass $M_{NS}\approx1.4M_{\odot}\approx1.57\times
10^{57}GeV$, a typical radius $R_{NS}\approx10km\approx5\times10^{19}GeV^{-1}%
$, with a neutron volume number density $n\approx\frac{M_{NS}/m}{(4/3)\pi
R_{NS}^{3}}\approx\pi\times10^{-3}GeV^{3}$, where $m$ is the neutron mass. (We
use natural units where, approximately, $1$fm $=5GeV^{-1}$, $1GeV^{-1}%
=6.6\times10^{-25}\sec=2.1\times10^{-31}$yr.) A slow moving dimension bubble
that gets gravitationally captured by a neutron star (NS) will be accelerated
to a high speed (near the escape speed for the NS) at its surface. Associated
with the high speed impact will be a large frictional force dissipating the
bubble's energy, and we will assume that the bubble quickly finds itself at
the center of the NS. The various simplifying assumptions are made to allow us
to extract some of the essential features characterizing the interaction
between the dimension bubble and the NS.

Now consider a type I dimension bubble of mass $M$ and radius $R$ sitting at
the center of the NS, spherically accreting the star's mass. We consider the
NS to be a degenerate Fermi gas with a volume number density of particles with
energy between $E$ and $E+dE$ given by $\mathcal{N}(E)dE$. Within a time $dt$
a volume $dV$ of a spherical annulus of the NS falls into the bubble at a rate
$\dot{V}=dV/dt$, carrying with it an amount of energy $dM$ which increases the
bubble mass by an amount $dM$ at a rate $\dot{M}=dM/dt$. We can then write the
rate of energy absorption by the bubble per unit of particle energy as%
\begin{equation}
\frac{d\dot{M}}{dE}=\mathcal{N}(E)E\dot{V} \label{e1}%
\end{equation}
\noindent For the radial infall of neutrons we have $\dot{V}=4\pi R^{2}v(E)$
with particle velocity $v(E)=p/E=(E^{2}-m^{2})^{1/2}/E$, so that%
\begin{equation}
\dot{V}=4\pi R^{2}\frac{(E^{2}-m^{2})^{1/2}}{E} \label{e2}%
\end{equation}
In the previous analysis one has to correct however by two effects: only half
of the particles move in the direction of the bubble (this gives a factor of
one half). Second, of those particles that move in the direction of the
bubble, they do so at an angle. Given that the average of the cosinus of the
angle between the normal and the velocity is also a half, eq.(\ref{e1}) is to
be corrected by a factor of $\frac{1}{4}$. The number density $\mathcal{N}(E)$ is%

\begin{equation}
\mathcal{N}(E)=2\cdot\frac{1}{2\pi^{2}}\frac{E(E^{2}-m^{2})^{1/2}}%
{e^{\beta(E-E_{F})}+1} \label{e3}%
\end{equation}
where the factor 2 represents the number of spin degrees of freedom per
particle and $E_{F}$ is the Fermi energy. We consider the low temperature
$\beta\rightarrow\infty$ limit where%
\begin{equation}
\frac{1}{e^{\beta(E-E_{F})}+1}\approx\left\{
\begin{array}
[c]{cc}%
1, & E<E_{F}\\
0, & E>E_{F}%
\end{array}
\right\}  \label{e4}%
\end{equation}
Integrating (\ref{e1}), corrected by the aboved mentioned factor of $\frac
{1}{4}$, and using (\ref{e2})-(\ref{e4}) we obtain%
\begin{equation}
\dot{M}=\int_{m}^{\infty}\mathcal{N}(E)E\dot{V}~dE=\frac{p_{F}^{4}R^{2}}{4\pi}
\label{e5}%
\end{equation}
where $p_{F}=(E_{F}^{2}-m^{2})^{1/2}$ is the Fermi momentum. The Fermi
momentum is related to the volume number density $n$ of neutrons by
$p_{F}=(3\pi^{2}n)^{1/3}$. Taking $n\approx\pi\times10^{-3}GeV^{3}$, we have
$p_{F}\approx.45~GeV$, $E_{F}\approx1.04~GeV\sim m$, and (\ref{e5}) gives%
\begin{equation}
\dot{M}\approx~8n^{4/3}R^{2}\approx(3.7\times10^{-3}~GeV^{4})R^{2} \label{e6}%
\end{equation}

The result in (\ref{e6}) can be compared to a rough estimate obtained from the
simple expression%
\begin{equation}
\dot{M}\sim\frac{1}{4}\rho_{NS}\dot{V}\sim\pi R^{2}\rho_{NS}v \label{e7}%
\end{equation}
where we take $\dot{V}\sim4\pi R^{2}v$ and again insert the above mentioned
factor of $\frac{1}{4}$. Using $\rho_{NS}\sim nE_{F}\sim mn$ for the average
energy density of particles in the neutron star and $v\sim p_{F}/E_{F}\sim.43$
for a typical velocity of a particle falling radially into the bubble, the
simple expression in (\ref{e7}) gives $\dot{M}\sim(4\times10^{-3}GeV^{4}%
)R^{2}$, which is very close to the result in (\ref{e6}).

\subsection{Quantum Reflection}

If the size of the extra dimension(s) is much larger inside the bubble than
outside, the mass of the particle is greatly reduced inside the bubble. In
effect, at the classical level, the particle experiences an enormous
attractive force toward the bubble's interior within the bubble wall. Quantum
mechanically, we can consider an associated attractive potential for the
bubble's interior due to the decrease in particle mass. But even an attractive
potential will generally give rise to some probability of reflection, with a
corresponding reflection coefficient $\mathcal{R}$. Here, we wish to get a
rough idea of how important this quantum reflection might be for particles
being absorbed by a dimension bubble.

To do so, we ignore particle spin and polarization effects and consider, for
simplicity, a one dimensional problem of relativistic spinless bosons obeying
the Klein-Gordon equation $(\square+m_{eff}^{2}(x))\phi=0$ scattering from a
flat bubble wall located at $x=0$, which separates the bubble
\textquotedblleft exterior\textquotedblright\ for $x<0$ and the bubble
\textquotedblleft interior\textquotedblright\ for $x>0$. We approximate the
position dependence of the boson mass with a step function, $m_{eff}%
^{2}(x)=m^{2}[1-\theta(x)]$. In the \textquotedblleft
exterior\textquotedblright\ region ($x<0$) we assume there to be a beam of
particles of mass $m$ incident upon the wall from the left, propagating in the
$+x$ direction, along with a reflected beam propagating in the $-x$ direction.
For the \textquotedblleft interior\textquotedblright\ region ($x>0$) there is
a beam of transmitted massless particles propagating in the $+x$ direction. We
then have the plane wave solutions%
\begin{equation}%
\begin{array}
[c]{lll}%
\phi_{ext}=\phi_{0}+\phi_{1} & =A_{0}e^{i(Et-px)}+A_{1}e^{i(Et+px)}, & (x<0)\\
\phi_{int}=\phi_{2} & =A_{2}e^{i(Et-p^{\prime}x)}, & (x>0)
\end{array}
\label{a2}%
\end{equation}
where $E^{2}=p^{2}+m^{2}=p^{\prime2}$. Demanding continuity of $\phi$ and
$\partial\phi/\partial x$ at $x=0$ leads to%
\begin{equation}
\frac{A_{1}}{A_{0}}=-\left(  \frac{E-p}{E+p}\right)  ,\ \ \ \ \ \frac{A_{2}%
}{A_{0}}=\frac{2p}{E+p} \label{a3}%
\end{equation}
We define the reflection coefficient $\mathcal{R}=j_{1}/j_{0}$ as the ratio of
the reflected current density $j_{1}=i\phi_{1}^{\ast}\overleftrightarrow
{\partial_{x}}\phi_{1}$ to the incident current density $j_{0}=i\phi_{0}%
^{\ast}\overleftrightarrow{\partial_{x}}\phi_{0}$, giving%
\begin{equation}
\mathcal{R}=\left(  \frac{E-p}{E+p}\right)  ^{2} \label{a4}%
\end{equation}
Similarly, the transmission coefficient is found to be $\mathcal{T}%
=j_{2}/j_{0}=4Ep/(E+p)^{2}$ with $\mathcal{R+T=}1$. Taking typical values
$p\sim p_{F}\sim m/2$, $E\sim E_{F}\sim2p_{F}\sim m$, we obtain $\mathcal{R}%
\sim1/9$, $\mathcal{T}\sim8/9$. Although the reflection probability may be
near 10 percent, we will neglect it in what follows.

\section{Near-Equilibrium Bubbles}

We \ again consider a type I dimension bubble where the size of the extra
dimension(s) inside the bubble is much greater than that outside the bubble,
i.e., $B_{in}\gg B_{out}$, where $B_{in(out)}$ denotes the scale factor of the
extra dimension(s). Massive particle modes then have masses $m_{in}\ll
m_{out}$, and we can focus on the effective radiation modes comprised of
photons as well as particles with masses $m_{in}\ll|\vec{p}|$ that tend to
stabilize the bubble against collapse. (We assume that the energy density of
any nonrelativistic massive modes is negligible in comparison to the radiation
energy density $\rho_{rad}\propto T^{4}$, where $T$ is the temperature inside
the bubble.) We also assume that the energy density $\lambda$ associated with
the value of the effective potential inside the bubble is negligible in
comparison to the radiation energy density $\rho_{rad}$. The mass $M$ of such
a bubble in mechanical equilibrium is given, approximately, in terms of the
bubble radius $R$ by\cite{GM,JM1}%

\begin{equation}
M=12\pi\sigma R^{2} \label{e8}%
\end{equation}
where $\sigma$ is the surface energy density of the bubble wall. If we
consider a bubble consuming matter that adjusts its radius sufficiently
quickly to be considered in a state of near-equilibrium, we can use (\ref{e8})
so that (\ref{e6}) gives%
\begin{equation}
\frac{\dot{M}}{M}\approx0.2\frac{n^{4/3}}{\sigma}=\frac{K}{\sigma_{GeV}}
\label{e9}%
\end{equation}
where $\sigma_{GeV}$ is a dimensionless number given in terms of $\sigma$ by
$\sigma=\sigma_{GeV}(GeV^{3})$ and%
\begin{equation}
K=0.2n^{4/3}\approx1\times10^{-4}GeV\approx1.5\times10^{20}\sec^{-1}%
\approx5\times10^{27}~\text{yr}^{-1} \label{e10}%
\end{equation}
On dimensional grounds, we take a typical time scale associated with bubble
equilibration to be $\tau_{eq}\sim\sigma^{-1/3}$, so that (\ref{e9}) is
assumed to be roughly self consistent for $(\dot{M}/M)\lesssim\tau_{eq}^{-1}$,
or $\sigma^{1/3}\gtrsim10^{-1}~GeV$. An integration of (\ref{e9}) indicates
that the bubble acquires a mass $M$ after a time%
\begin{equation}
t=\frac{\sigma_{GeV}}{K}\ln\frac{M}{M_{0}} \label{e11}%
\end{equation}
where $M_{0}$ is the initial bubble mass.

\section{Black Hole Formation, Core Collapse}

We must consider the possibility that a growing dimension bubble will form a
black hole before consuming the entire NS. For a spherical nonrotating bubble
this occurs when the bubble mass reaches a value $M_{crit}$ such that
$R=2GM_{crit}$. Using this in conjunction with (\ref{e8}) gives%
\begin{equation}
M_{crit}=\frac{M_{P}^{4}}{48\pi\sigma},\ \ \ \ \ \frac{M_{crit}}{M_{NS}}%
\sim\frac{4.2\times10^{16}}{\sigma_{GeV}} \label{e12}%
\end{equation}
where $M_{P}=G^{-1/2}\sim10^{19}GeV$ is the Planck mass. From (\ref{e11}) the
time required for the bubble to evolve into a black hole is%
\begin{equation}
t_{crit}=(\sigma_{GeV}/K)\ln(M_{crit}/M_{0}), \label{e13}%
\end{equation}
provided that $M_{crit}\lesssim M_{NS}$, or $\sigma^{1/3}\gtrsim
3.4\times10^{5}~GeV$. For $\sigma^{1/3}\lesssim3.4\times10^{5}~GeV$, the
bubble can devour the entire NS within a time%
\begin{equation}
\tau\sim(\sigma_{GeV}/K)\ln(M_{NS}/M_{0}) \label{e14}%
\end{equation}
without first forming a black hole. (The logarithmic functions $\ln
(M_{crit}/M_{0})$ and $\ln(M_{NS}/M_{0})$ are slowly varying functions of
$M_{0}$, with $\ln(M_{NS}/M_{0})\lesssim130$ for $M_{0}\gtrsim1\ GeV$, and are
therefore relatively insensitive to $M_{0}$.)

Table \ref{t1} \ presents some values (with $M_{0}=1GeV$) indicating
$M_{crit}$, along with $\tau$ (amount of time required for bubble to consume
entire NS without formation of a black hole) or $t_{crit}$ (amount of time
required for bubble to grow within NS before formation of a black hole) for
various values of the bubble energy scale $\sigma_{GeV}^{1/3}$.

\bigskip%

\begin{table}[htbp] \centering
\begin{tabular}
[t]{|c|c|c|c|}\hline
$\mathbf{\sigma}_{GeV}^{1/3}$ & $\dfrac{M_{crit}}{M_{NS}}$ & $\tau$ (yr) &
$\dfrac{t_{crit}\text{(yr)}}{\ln(M_{crit}/M_{0})}$\\\hline
10$^{-1}$ & 4.2$\times$10$^{19}$ & 3$\times$10$^{-29}$ & --\\\hline
10$^{0}$ & 4.2$\times$10$^{16}$ & 3$\times$10$^{-26}$ & --\\\hline
10$^{3}$ & 4.2$\times$10$^{7}$ & 3$\times$10$^{-17}$ & --\\\hline
10$^{6}$ & 4.2$\times$10$^{-2}$ & -- & 2.2$\times$10$^{-10}$\\\hline
10$^{9}$ & 4.2$\times$10$^{-11}$ & -- & 2.2$\times$10$^{-1}$\\\hline
10$^{12}$ & 4.2$\times$10$^{-20}$ & -- & 2.2$\times$10$^{8}$\\\hline
10$^{16}$ & 4.2$\times$10$^{-32}$ & -- & 2.2$\times$10$^{20}$\\\hline
10$^{19}$ & 4.2$\times$10$^{-41}$ & -- & 2.2$\times$10$^{29}$\\\hline
\end{tabular}
\caption{Some characteristics associated with a type I dimension bubble inside a
neutron star }\label{t1}%
\end{table}%

If we use the Bondi-Hoyle accretion formula\cite{BHL}%
\begin{equation}
\dot{M}=4\pi G^{2}\rho_{NS}M^{2} \label{e15}%
\end{equation}
to describe the spherical accretion of the NS remainder by the black hole, we
find that the amount of time $t_{BH}$ required for the black hole mass to
evolve from $M_{crit}$ to $M_{NS}$, i.e., the amount of time needed for the
black hole to consume the entire NS, is%
\begin{equation}
t_{BH}=(4\pi G^{2}\rho_{NS})^{-1}\left(  \frac{1}{M_{crit}}-\frac{1}{M_{NS}%
}\right)  \label{e16}%
\end{equation}
which may vary from $t_{BH}\sim10^{-3}\sec$ for $\sigma_{GeV}^{1/3}=10^{6}$ to
$t_{BH}\sim10^{8}$yr for $\sigma_{GeV}^{1/3}=10^{12}$ to $t_{BH}\sim10^{15}$yr
for $\sigma_{GeV}^{1/3}=10^{19}$.

It is possible that a neutron star will become destabilized by an evolving
dimension bubble or black hole and a core collapse of the NS may lead to an
observable photon burst from the collapsing/exploding star. However, the
ultimate astrophysical outcome and phenomenology will be highly dependent upon
the particle physics model parameters such as $\sigma$, as well as the
astrophysics connected with the NS collapse and the potentially ensuing explosion.

\

\end{document}